\documentclass[%
reprint,
superscriptaddress,
showpacs,
preprintnumbers,
 bibnotes,
 amsmath,amssymb,
 aps,
prl
]{revtex4-1}

\usepackage{graphicx}% Include figure files
\usepackage{dcolumn}% Align table columns on decimal point
\usepackage{bm}% bold math
\usepackage{color}
\begin{document}

%\preprint{Preprint}

\title{Prominent role of spin-orbit coupling in FeSe}

       %Effects of  Residual Stress and Uniaxial Strain  on the Observation of  Nemacticity in    $\mathrm{BaFe_2As_2}$ by Raman Scattering
%}
\author{Mingwei~Ma}
\affiliation{International Center for Quantum Materials, School of Physics, Peking University, Beijing 100871, China}
\author{Philippe~Bourges}
\affiliation{Laboratoire L\'{e}on Brillouin, CEA-CNRS, Universit\'{e} Paris-Saclay, CEA Saclay, 91191 Gif-sur-Yvette, France}
\author{Yvan~Sidis}
\affiliation{Laboratoire L\'{e}on Brillouin, CEA-CNRS, Universit\'{e} Paris-Saclay, CEA Saclay, 91191 Gif-sur-Yvette, France}
\author{Yang~Xu}
\affiliation{State Key Laboratory of Surface Physics, Department of Physics, and Laboratory of Advanced Materials, Fudan University, Shanghai 200433, China}
\author{Shiyan~Li}
\affiliation{State Key Laboratory of Surface Physics, Department of Physics, and Laboratory of Advanced Materials, Fudan University, Shanghai 200433, China}
\affiliation{Collaborative Innovation Center of Advanced Microstructures, Nanjing 210093, China}
\author{Biaoyan~Hu}
\affiliation{International Center for Quantum Materials, School of Physics, Peking University, Beijing 100871, China}
\author{Jiarui~Li}
\affiliation{International Center for Quantum Materials, School of Physics, Peking University, Beijing 100871, China}
\author{Fa~Wang}
\affiliation{International Center for Quantum Materials, School of Physics, Peking University, Beijing 100871, China}
\affiliation{Collaborative Innovation Center of Quantum Matter, Beijing 100871, China}
\author{Yuan~Li}
\email[]{yuan.li@pku.edu.cn}
\affiliation{International Center for Quantum Materials, School of Physics, Peking University, Beijing 100871, China}
\affiliation{Collaborative Innovation Center of Quantum Matter, Beijing 100871, China}

\begin{abstract}
In most existing theories for iron-based superconductors, spin-orbit coupling (SOC) has been assumed insignificant. Even though recent experiments have revealed an influence of SOC on the electronic band structure, whether SOC fundamentally affects magnetism and superconductivity remains an open question. Here we use spin-polarised inelastic neutron scattering to show that collective low-energy spin fluctuations in the orthorhombic (or ``nematic") phase of FeSe possess nearly no in-plane component. Such spin-space anisotropy can only be caused by SOC. It is present over an energy range greater than the superconducting gap 2$\Delta _\mathrm{sc}$ and gets fully inherited in the superconducting state, resulting in a distinct $c$-axis polarised ``spin resonance". Our result demonstrates the importance of SOC in defining the low-energy spin excitations in FeSe, which helps to elucidate the nearby magnetic instabilities and the debated interplay between spin and orbital degrees of freedom. The prominent role of SOC also implies a possible unusual nature of the superconducting state.
\end{abstract}

%\pacs{74.70.Xa, %Pnictides and chalcogenides
%75.30.Gw, %Magnetic anisotropy
%71.70.Ej, %Spin-orbit coupling, Zeeman and Stark splitting, Jahn-Teller effect
%78.70.Nx  %Neutron inelastic scattering
%}

\maketitle
Spin-orbit coupling (SOC) is a fundamental interaction in solids due to the relativistic motion of electrons. The effect of SOC is expected to be pronounced in the presence of heavy elements, whereas its relevance to the iron-based superconductors has thus far been assumed insignificant in most theoretical treatment. This assumption has led to a separation of ideas, in particular those attempting to explain the formation of superconductivity \cite{HirschfeldRepProgPhys2011,ScalapinoRevModPhys2012,KontaniPhysRevLett2010} and nematic order \cite{QisiWangNatMater2016,FernandesNatPhys2014,BaekNatMater2016,WangFaNatPhys2015,GlasbrenerNatPhys2015,YuRongPhysRevLett2015,YamakawaPhysRevX2016} into ``spin" and ``orbital" camps. Indeed, in the presence of SOC, neither the spin nor the orbital angular momentum remains a good quantum number, hence any collective electronic behaviour would have to be considered as a consequence of the joint interactions involving both.

Recent angle-resolved photoemission spectroscopy (ARPES) experiments have demonstrated the presence of SOC in iron-based superconductors, via the observation of electronic band splitting at the Brillouin zone (BZ) center in the tetragonal phase \cite{WatsonPhysRevB2015,JohnsonPhysRevLett2015,BorisenkoNatPhys2015}. The affected quasiparticle states can be expected to have spin structures that will give rise to spin-space anisotropy (SSA) in the magnetism, if the essence of the magnetism can be captured by an itinerant (spin-density-wave-like) description \cite{DaiRevModPhys2015}. Indeed, an energy gap related to SSA has been observed by inelastic neutron scattering (INS) in pnictide parent compounds \cite{ParKPhysRevB2012} with long-range stripe-antiferromagnetic order. Spin-polarised INS experiments \cite{QureshiPhysRevB2012,WangCPhysRevX2013} have further shown that the SSA is ``$XYZ$" (orthorhombic) rather than ``$XXZ$" (tetragonal), which suggests that the orbital structure in the nematic phase might be its origin \cite{QureshiPhysRevB2012}. However, since the nematic order is closely accompanied by magnetic order in the pnictides, it remains unclear whether the SSA depends on the presence of the nearby magnetic order. Moreover, much of the INS results obtained in the magnetically ordered state have only been discussed under a local-moment description of the magnetism with phenomenological single-ion anisotropy \cite{QureshiPhysRevB2012,WangCPhysRevX2013}. In order to assess the influence of SOC on the magnetism at a more fundamental level, in particular using an itinerant description in conjunction with the ARPES results, it is highly desirable to study SSA in a system that has no magnetic order.

FeSe offers an ideal opportunity for this purpose, since it has nematic order below $T_\mathrm{s}$ $\approx$ 90 K but no magnetic order down to the lowest temperature \cite{McQueenPhysRevLett2009}. Figure 1a displays the crystal and Fermi surface (FS) structure of FeSe in the nematic phase \cite{YeZR2015}, along with leading Fe 3$d$ orbital characters ($xy$, $xz$ and $yz$) of quasiparticle states on the FS. Throughout our presentation, we use the 4-Fe unit cell, for which the BZ is shown by the dashed lines in Fig. 1a. The centers of the hole and electron pockets are connected by two-dimensional wave vectors (1, 0) and (0, 1) in reciprocal lattice units (r.l.u.).

\begin{figure}
\includegraphics[width=3.375in]{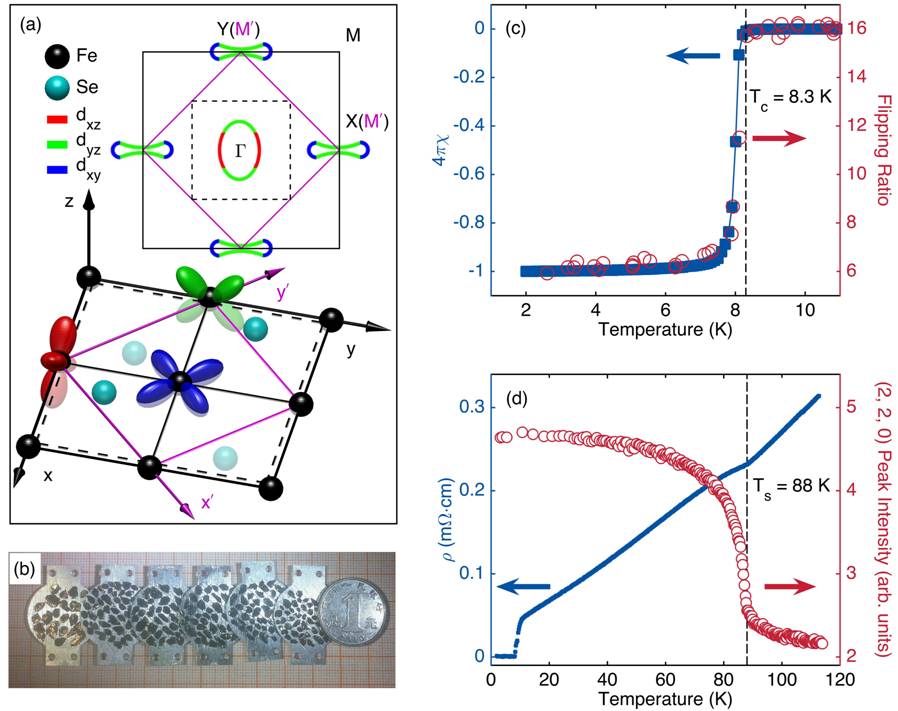}
\caption{\label{fig1}
Crystal structure, FS topology, and characterisation of phase transition temperatures in FeSe single crystals. (a) Crystal structure and FS topology \cite{YeZR2015}, with orbital characters color-coded with the orbital shapes. Three Brillouin-zone notations are shown for comparison: 1-Fe unit cell (black solid lines), 2-Fe unit cell (magenta solid lines), and 4-Fe unit cell (dashed lines, used throughout our presentation). (b) Photograph of our INS sample before the final assembly. (c) DC magnetic susceptibility, $\chi$, shows a sharp superconducting transition at $T_\mathrm{c}$ = 8.3 K in zero-field-cooled measurement in a magnetic field of H = 10 Oe applied parallel to the $ab$ plane. A consistent transition is observed for the entire INS sample using depolarisation effect on the neutron beam. (d) Resistivity, $\rho$, shows an abrupt change in slope vs. $T$ at the nematic transition temperature $T_\mathrm{s}$ = 88 K, below which an abrupt increase in the (2, 2, 0) Bragg scattering intensity is observed due to reduced extinction effects.}
\label{fig1}
\end{figure}

Our spin-polarised INS experiment was carried out on the 4F1 triple-axis spectrometer at the Laboratoire L\'{e}on Brillouin, Saclay, France (the instrument configurations are described in the Supplementary Information). Over four hundred single crystals of FeSe (Fig. 1b) were used for this study, with a total mass of about 3.5 grams. They were grown with a chemical-vapour-transport method and co-aligned within 6$^\circ$ mosaicity in the ($H$, $K$, 0) scattering plane on aluminium plates using a hydrogen-free adhesive. Due to the heavy loss in incident flux and detection efficiency associated with current polarised-neutron techniques, very long measurement time (3 hours or more per data point) was required to acquire satisfactory statistical accuracy. The quality of spin polarisation in a neutron scattering experiment can be quantified by the flipping ratio, defined here as the ratio between intensities measured in non-spin-flip and spin-flip geometries on strong nuclear Bragg peaks. The flipping ratio in our experiment is about 16 for all neutron spin polarisations (Fig. 1c). Because the spin polarisation is maintained by a guide magnetic field ($\sim10$ Oe) throughout the beam path, the beam can be partially depolarised if the guide field changes abruptly, such as at the sample surface when the sample is in a diamagnetic (Meissner's) state below $T_\mathrm{c}$. This offers a method to measure $T_\mathrm{c}$ for the entire sample array. To do this, the sample was cooled below $T_\mathrm{c}$ in a guide field so that it contained trapped vortices, and then the guild field direction was rotated by 90$^\circ$. The flipping ratio was reduced to $\sim6$ after the guide-field rotation and was measured upon warming up the sample (Fig. 1c), with a recovery to its original value at $T_\mathrm{c}$. The superconducting critical temperature, $T_\mathrm{c}$ $\approx$ 8.3 K (Fig. 1c), and the nematic transition temperature, $T_\mathrm{s}$ $\approx$ 88 K (Fig. 1d), were determined on selected crystals by magnetic susceptibility and resistivity measurements, respectively, as well as on the entire sample by neutron scattering methods.

\begin{figure}
\includegraphics[width=3.375in]{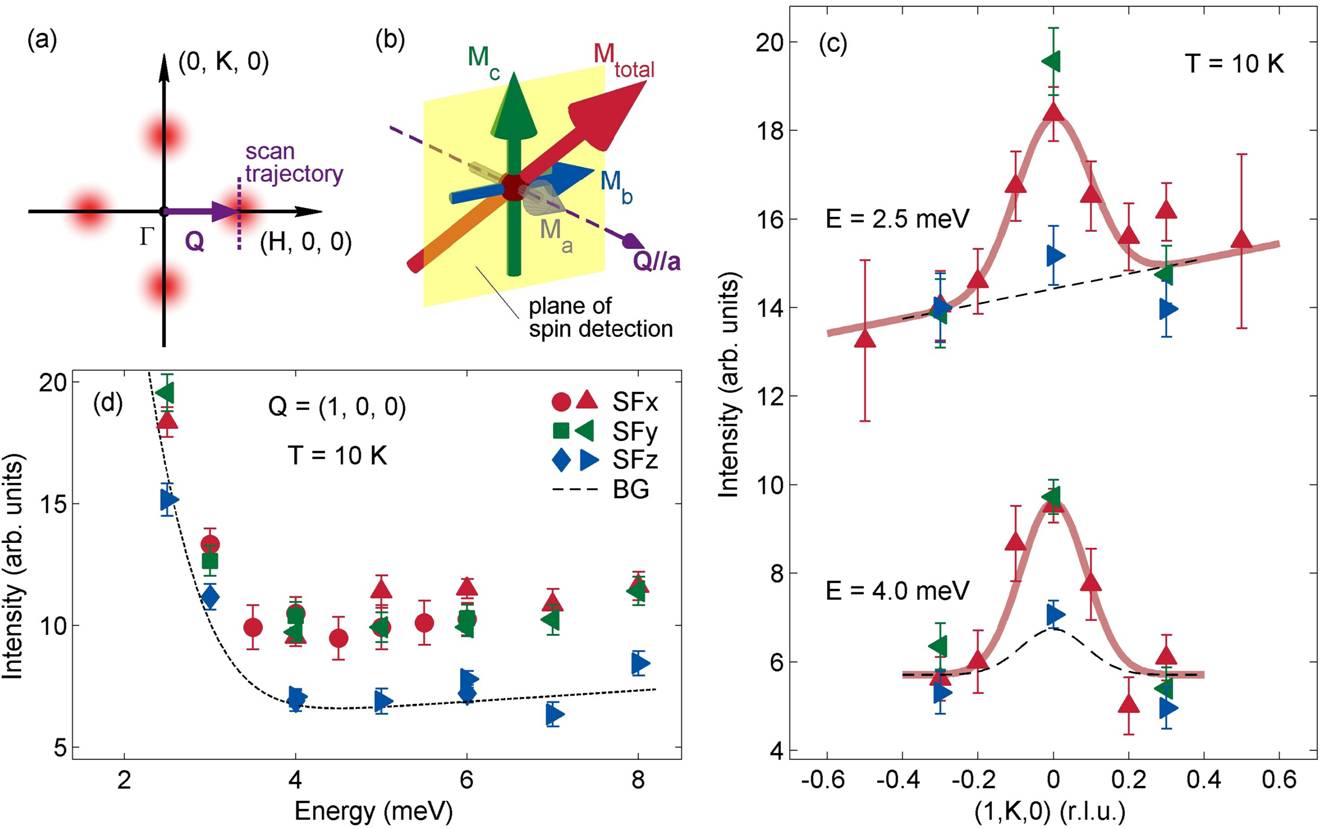}
\caption{\label{fig2}
Spin-polarised INS selection rules and predominant $M_{c}$ contribution to the magnetic signal. (a) Distribution of low-energy magnetic signals in momentum space. (b) Total spin fluctuations ($M_\mathrm{total}$) and its three components, two of which can be detected by INS at \textbf{Q} = (1, 0, 0) with selection rules described in the text. (c) Momentum scans at fixed energies ($E$) along a trajectory shown in (a). (d) Energy scans at fixed \textbf{Q} = (1, 0, 0). (c) and (d) share the same legends that indicate the three spin-flip channels. The two sets of symbols in (d) represent data obtained with different sample-environment devices. The BG intensity is determined from the selection rule: BG = SF$_{b}$ + SF$_{c}$ $-$ SF$_\mathrm{\textbf{Q}}$, taking all available data points into account (see Supplementary Information).}
\label{fig2}
\end{figure}

Previous unpolarised INS studies on FeSe \cite{QisiWangNatMater2016,RahnPhysRevB2015}, have uncovered strong magnetic signals at \textbf{Q} = (1, 0, 0) below $T_\mathrm{s}$ and a spin resonance at 4 meV below $T_\mathrm{c}$. For INS samples which are typically twinned, the same scattering signal can be expected at \textbf{Q} = (0, 1, 0), as depicted in Fig. 2a. For convenience, we keep the notation \textbf{Q} = (1, 0, 0), and the nominal $b$ direction should be understood as within the FeSe plane and perpendicular to Q. As illustrated in Fig. 2b, INS only detects spin fluctuations in directions perpendicular to \textbf{Q}, $i.e.$, along $b$ ($M_{b}$) and $c$ ($M_{c}$) for \textbf{Q} = (1, 0, 0). By analysing the neutron-spin dependence of the scattering signal, we are able to determine these two components separately. Most of our data were obtained in spin-flip channels with the incoming neutron spin along \textbf{Q} (SF$_\mathrm{\textbf{Q}}$), $b$ (SF$_{b}$), and $c$ (SF$_{c}$) directions, which detect the $M_{b}$ + $M_{c}$, $M_{c}$, and $M_{b}$ components, respectively, on top of a common background (BG).

Figure 2c displays constant-energy scans at 2.5 and 4.0 meV, performed at 10 K along a trajectory shown in Fig. 2a. A clear commensurate peak is seen in the SF$_\mathrm{\textbf{Q}}$ channel, consistent with previous unpolarised INS results \cite{QisiWangNatMater2016}. If the scattering is isotropic in spin space, the signal should be equally distributed in the SF$_{b}$ and SF$_{c}$ channels, which is clearly not the case. By comparing data obtained in the three different channels, we are able to determine the BG intensity (see Supplementary Information), which we plot in Fig. 2d together with raw data from energy scans measured at fixed \textbf{Q} = (1, 0, 0). We find that the intensities in the SF$_\mathrm{\textbf{Q}}$ and SF$_{b}$ channels are nearly equal over the entire measured energy range 2.5 meV $\le$ E $\le$ 8 meV. This means that the SF$_{c}$ intensity is mostly BG, and that the magnetic signal is dominated by its $M_{c}$ component.

By measuring only at \textbf{Q} = (1, 0, 0), we are nominally not sensitive to the $M_{a}$ component (Fig. 2b). For a twinned sample below $T_\mathrm{s}$, however, we simultaneously detect magnetic signals from the two nematic domains, which is equivalent to detecting physical signals from both \textbf{Q}$_1$ = (1, 0, 0) and \textbf{Q}$_2$ = (0, 1, 0), with \{$M_{b}$, $M_{c}$\} and \{$M_{a}$, $M_{c}$\} components, respectively. Our data show that $M_{b}$(\textbf{Q}$_1$) + $M_{a}$(\textbf{Q}$_2$) is negligible compared to $M_{c}$(\textbf{Q}$_1$) + $M_{c}$(\textbf{Q}$_2$). If low-energy spin excitations are controlled by FS nesting \cite{DaiRevModPhys2015}, we believe that the overall intensity difference between \textbf{Q}$_1$ and \textbf{Q}$_2$ is not large, because the orbital characters and quality of FS nesting are very similar (Fig. 1a). Our result hence indicates that the low-energy spin excitations have nearly no in-plane components ($M_{a}$ and $M_{b}$).

We present in Fig. 3 the evolution of spin fluctuations with temperature. In the tetragonal phase, the overall intensity is weak and the data are consistent with an isotropic distribution of scattering signals in $M_{b}$ and $M_{c}$. Upon cooling into the nematic phase, a strong enhancement is found in $M_{c}$ but not in $M_{b}$. Upon further cooling below $T_\mathrm{c}$, the magnetic spectral weight rearranges itself into a spin resonance \cite{QisiWangNatMater2016} at 4 meV. Our data unambiguously show that the resonance is essentially fully $c$-axis polarised (Fig. 3d).

\begin{figure}
\includegraphics[width=2.8in]{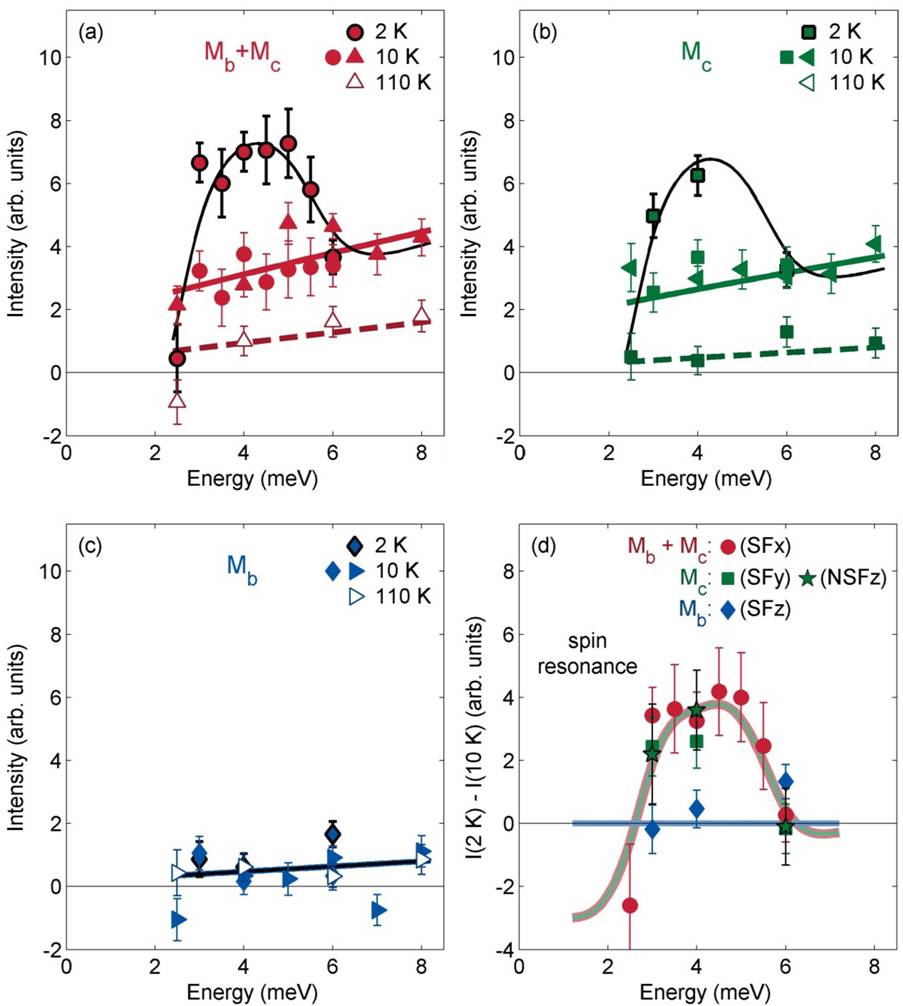}
\caption{\label{fig3}
Evolution of different spin-fluctuation components with temperature. (a)$-$(c) Net magnetic signal components at \textbf{Q} = (1, 0, 0) at three different temperatures, obtained by subtracting the globally determined BG intensity (see Fig. 2 and Supplementary Information) from SF$_\mathrm{\textbf{Q}}$, SF$_{b}$ and SF$_{c}$ data, respectively. (d) Intensity change across $T_\mathrm{c}$ measured in different geometries. The non-spin-flip scattering geometry with incoming neutron spin polarisation along $c$ (NSF$_{c}$) measures spin fluctuations along the $c$ direction and confirms the spin-flip (SF$_{b}$) result. Solid lines are guide to the eye. The two sets of symbols in (a)$-$(c) for 10 K represent data obtained with different sample-environment devices (see Supplementary Information).
}
\label{fig3}
\end{figure}

It is revealing to compare our results to SSA observed in other unconventional superconductors. We begin by noting that spin excitations at the lowest energies in magnetically ordered pnictides \cite{QureshiPhysRevB2012,WangCPhysRevX2013} are dominated by $M_{c}$, because $M_{c}$ has a smaller energy gap than $M_{b}$ and $M_{a}$. Somewhat reduced but non-zero SSA has been observed on the spin resonance in doped pnictides \cite{SteffensPhysRevLett2013,ZhangPhysRevB2013}, also with larger spectral weight in $M_{c}$ than in $M_{b}$. Our observation of qualitatively similar but more pronounced SSA in paramagnetic FeSe suggests that, contrary to previous conjectures \cite{SteffensPhysRevLett2013,ZhangPhysRevB2013}, the SSA is unrelated to the stripe-antiferromagnetic order, but directly follows from the orbital structure associated with the nematic phase. The fact that FeSe exhibits the most clear-cut SSA on the spin resonance is likely a joint consequence of strong SOC and the unique electronic structure of FeSe: both the Fermi energy \cite{KasaharaPNAS2014} and the superconducting gap 2$\Delta _\mathrm{sc}$ are small \cite{LinPhysRevB2011,KasaharaPNAS2014,BorisenkoNatPhys2015}, compared to 8 meV up to which we are able to detect the SSA.

The observed SSA can be qualitatively understood using an itinerant description of the magnetism, with reasoning similar to that for Sr$_2$RuO$_4$. Low-energy spin fluctuations in Sr$_2$RuO$_4$ also have a leading $c$-axis component \cite{BradenPhysRevLett2004}, which can be attributed to the ruthenium $d_{xz}$ and $d_{yz}$ orbital character of quasiparticle states that are most involved in FS nesting \cite{EreminPhysRevB2002}. The same applies to FeSe: the quasiparticle states closest to the Fermi level near the $\mathrm\Gamma$ and M' points (Fig. 1a) are of predominant Fe $d_{xz}$ and $d_{yz}$ orbital character \cite{YeZR2015,Fedorov2016}. In the limits of pure $d_{xz}$/$d_{yz}$ orbital character and strong atomic SOC, the low-energy electronic states around both the hole and electron pockets will be  $\mid$$xz + i\cdot yz$,$\uparrow$$>$  and $\mid$$xz - i\cdot yz$,$\downarrow$$>$. The matrix elements of in-plane spin components between these states vanish (see Supplementary Information). For both FeSe and the pnictides \cite{QureshiPhysRevB2012}, it remains a theoretical challenge to explain the fact that SSA becomes more pronounced in the nematic state, in which the splitting between $d_{xz}$ and $d_{yz}$ orbitals is generally expected to weaken the above spin-orbital entanglement.

Our result has important implications on the magnetism in iron-based superconductors. Characteristics of low-energy spin excitations are usually linked to the nature of nearby magnetic instabilities. While spin fluctuations near (1, 0) indicate the presence of magnetic interactions in FeSe that are in favour of stripe-antiferromagnetism \cite{QisiWangNatMater2016,RahnPhysRevB2015}, the fact that these fluctuations are predominantly $c$-axis oriented implies that the leading magnetic instability would result in moments along the $c$-axis, consistent with a recent observation under pressure \cite{Wang2016}. Moreover, since the observed SSA can be qualitatively explained by the orbital structure, which is to a large extent ubiquitous to all iron-based superconductors, our result is consistent with the notion that the spin-reorientation transition in the pnictides \cite{WassPhysRevB2015,AllredNatPhys2016} arises from a competing magnetic instability that requires the presence of SOC \cite{ChristensenPhysRevB2015}.

The strong SOC not only helps to reconcile the debate on the spin-orbital interplay, $e.g.$, in driving the nematic order in FeSe \cite{FernandesNatPhys2014,BaekNatMater2016,WangFaNatPhys2015,GlasbrenerNatPhys2015,YuRongPhysRevLett2015,YamakawaPhysRevX2016}, but might also give rise to novel superconductivity \cite{YuanPhysRevLett2014,LuScience2015} by mixing the spin-singlet and triplet Cooper-pairing channels. A consequence of such mixture is that Cooper pairs become more robust against applied magnetic fields \cite{LuScience2015}. We confirm an earlier report \cite{HerSupercondSciTech2011} of in-plane upper critical field $H_\mathrm{c2}$ in FeSe as large as 27 Tesla (see Supplementary Information), which exceeds or approaches the Pauli limit $H_\mathrm{p}$ = 1.414 $\Delta _\mathrm{sc}$/g $\mathrm\mu$$_\mathrm{B}$ ($g$: Land\'{e} $g$ factor, taken to be 2 here; $\mathrm\mu$$_\mathrm{B}$: Bohr magneton) that amounts to 15.9 $-$ 30.5 Tesla for reported $\Delta _\mathrm{sc}$ values ranging from 1.3 to 2.5 meV \cite{LinPhysRevB2011,KasaharaPNAS2014,BorisenkoNatPhys2015}. It might not be a coincidence that a robust zero-energy bound state \cite{YinNatPhys2015}, suggestive of topological superconductivity, was recently observed in Fe(Te,Se) which has even stronger SOC than FeSe \cite{WatsonPhysRevB2015,JohnsonPhysRevLett2015,BorisenkoNatPhys2015}.

\begin{acknowledgments}

We wish to thank Jitae Park, Yan Zhang, Weiqiang Yu, Tao Li, Gang Chen, Pengcheng Dai and Haihu Wen for discussions. This work is supported by the National Natural Science Foundation of China (Grants No. 11374024 and No. 11522429) and Ministry of Science and Technology of China (Grants No. 2015CB921302 and No. 2013CB921903). The work of FW was initiated at the Aspen Center for Physics which is supported by National Science Foundation grant PHY-1066293, and partially supported by a grant from the Simons Foundation.
\end{acknowledgments}

%\preprint{Preprint}

\bibliography{FeSe_Reference}% Produces the bibliography via BibTeX.

\begin{thebibliography}{35}%
\makeatletter
\providecommand \@ifxundefined [1]{%
 \@ifx{#1\undefined}
}%
\providecommand \@ifnum [1]{%
 \ifnum #1\expandafter \@firstoftwo
 \else \expandafter \@secondoftwo
 \fi
}%
\providecommand \@ifx [1]{%
 \ifx #1\expandafter \@firstoftwo
 \else \expandafter \@secondoftwo
 \fi
}%
\providecommand \natexlab [1]{#1}%
\providecommand \enquote  [1]{``#1''}%
\providecommand \bibnamefont  [1]{#1}%
\providecommand \bibfnamefont [1]{#1}%
\providecommand \citenamefont [1]{#1}%
\providecommand \href@noop [0]{\@secondoftwo}%
\providecommand \href [0]{\begingroup \@sanitize@url \@href}%
\providecommand \@href[1]{\@@startlink{#1}\@@href}%
\providecommand \@@href[1]{\endgroup#1\@@endlink}%
\providecommand \@sanitize@url [0]{\catcode `\\12\catcode `\$12\catcode
  `\&12\catcode `\#12\catcode `\^12\catcode `\_12\catcode `\%12\relax}%
\providecommand \@@startlink[1]{}%
\providecommand \@@endlink[0]{}%
\providecommand \url  [0]{\begingroup\@sanitize@url \@url }%
\providecommand \@url [1]{\endgroup\@href {#1}{\urlprefix }}%
\providecommand \urlprefix  [0]{URL }%
\providecommand \Eprint [0]{\href }%
\providecommand \doibase [0]{http://dx.doi.org/}%
\providecommand \selectlanguage [0]{\@gobble}%
\providecommand \bibinfo  [0]{\@secondoftwo}%
\providecommand \bibfield  [0]{\@secondoftwo}%
\providecommand \translation [1]{[#1]}%
\providecommand \BibitemOpen [0]{}%
\providecommand \bibitemStop [0]{}%
\providecommand \bibitemNoStop [0]{.\EOS\space}%
\providecommand \EOS [0]{\spacefactor3000\relax}%
\providecommand \BibitemShut  [1]{\csname bibitem#1\endcsname}%
\let\auto@bib@innerbib\@empty
%</preamble>
\bibitem [{\citenamefont {Hirschfeld}\ \emph {et~al.}(2011)\citenamefont
  {Hirschfeld}, \citenamefont {Korshunov},\ and\ \citenamefont
  {Mazin}}]{HirschfeldRepProgPhys2011}%
  \BibitemOpen
  \bibfield  {author} {\bibinfo {author} {\bibfnamefont {P.~J.}\ \bibnamefont
  {Hirschfeld}}, \bibinfo {author} {\bibfnamefont {M.~M.}\ \bibnamefont
  {Korshunov}}, \ and\ \bibinfo {author} {\bibfnamefont {I.~I.}\ \bibnamefont
  {Mazin}},\ }\href {\doibase http://dx.doi.org/10.1088/0034-4885/74/12/124508}
  {\bibfield  {journal} {\bibinfo  {journal} {Rep. Prog. Phys.}\ }\textbf
  {\bibinfo {volume} {74}},\ \bibinfo {pages} {124508} (\bibinfo {year}
  {2011})}\BibitemShut {NoStop}%
\bibitem [{\citenamefont {Scalapino}(2012)}]{ScalapinoRevModPhys2012}%
  \BibitemOpen
  \bibfield  {author} {\bibinfo {author} {\bibfnamefont {D.~J.}\ \bibnamefont
  {Scalapino}},\ }\href {\doibase 10.1103/RevModPhys.84.1383} {\bibfield
  {journal} {\bibinfo  {journal} {Rev. Mod. Phys.}\ }\textbf {\bibinfo {volume}
  {84}},\ \bibinfo {pages} {1383} (\bibinfo {year} {2012})}\BibitemShut
  {NoStop}%
\bibitem [{\citenamefont {Kontani}\ and\ \citenamefont
  {Onari}(2010)}]{KontaniPhysRevLett2010}%
  \BibitemOpen
  \bibfield  {author} {\bibinfo {author} {\bibfnamefont {H.}~\bibnamefont
  {Kontani}}\ and\ \bibinfo {author} {\bibfnamefont {S.}~\bibnamefont
  {Onari}},\ }\href {\doibase 10.1103/PhysRevLett.104.157001} {\bibfield
  {journal} {\bibinfo  {journal} {Phys. Rev. Lett.}\ }\textbf {\bibinfo
  {volume} {104}},\ \bibinfo {pages} {157001} (\bibinfo {year}
  {2010})}\BibitemShut {NoStop}%
\bibitem [{\citenamefont {Wang}\ \emph
  {et~al.}(2016{\natexlab{a}})\citenamefont {Wang}, \citenamefont {Shen},
  \citenamefont {Pan}, \citenamefont {Hao}, \citenamefont {Ma}, \citenamefont
  {Zhou}, \citenamefont {Steffens}, \citenamefont {Schmalzl}, \citenamefont
  {Forrest}, \citenamefont {Abdel-Hafiez}, \citenamefont {Chen}, \citenamefont
  {Chareev}, \citenamefont {Vasiliev}, \citenamefont {Bourges}, \citenamefont
  {Sidis}, \citenamefont {Cao},\ and\ \citenamefont
  {Zhao}}]{QisiWangNatMater2016}%
  \BibitemOpen
  \bibfield  {author} {\bibinfo {author} {\bibfnamefont {Q.}~\bibnamefont
  {Wang}}, \bibinfo {author} {\bibfnamefont {Y.}~\bibnamefont {Shen}}, \bibinfo
  {author} {\bibfnamefont {B.}~\bibnamefont {Pan}}, \bibinfo {author}
  {\bibfnamefont {Y.}~\bibnamefont {Hao}}, \bibinfo {author} {\bibfnamefont
  {M.}~\bibnamefont {Ma}}, \bibinfo {author} {\bibfnamefont {F.}~\bibnamefont
  {Zhou}}, \bibinfo {author} {\bibfnamefont {P.}~\bibnamefont {Steffens}},
  \bibinfo {author} {\bibfnamefont {K.}~\bibnamefont {Schmalzl}}, \bibinfo
  {author} {\bibfnamefont {T.~R.}\ \bibnamefont {Forrest}}, \bibinfo {author}
  {\bibfnamefont {M.}~\bibnamefont {Abdel-Hafiez}}, \bibinfo {author}
  {\bibfnamefont {X.}~\bibnamefont {Chen}}, \bibinfo {author} {\bibfnamefont
  {D.~A.}\ \bibnamefont {Chareev}}, \bibinfo {author} {\bibfnamefont {A.~N.}\
  \bibnamefont {Vasiliev}}, \bibinfo {author} {\bibfnamefont {P.}~\bibnamefont
  {Bourges}}, \bibinfo {author} {\bibfnamefont {Y.}~\bibnamefont {Sidis}},
  \bibinfo {author} {\bibfnamefont {H.}~\bibnamefont {Cao}}, \ and\ \bibinfo
  {author} {\bibfnamefont {J.}~\bibnamefont {Zhao}},\ }\href {\doibase
  10.1038/nmat4492} {\bibfield  {journal} {\bibinfo  {journal} {Nature Mater.}\
  }\textbf {\bibinfo {volume} {15}},\ \bibinfo {pages} {159} (\bibinfo {year}
  {2016}{\natexlab{a}})}\BibitemShut {NoStop}%
\bibitem [{\citenamefont {Fernandes}\ \emph {et~al.}(2014)\citenamefont
  {Fernandes}, \citenamefont {Chubukov},\ and\ \citenamefont
  {Schmalian}}]{FernandesNatPhys2014}%
  \BibitemOpen
  \bibfield  {author} {\bibinfo {author} {\bibfnamefont {R.~M.}\ \bibnamefont
  {Fernandes}}, \bibinfo {author} {\bibfnamefont {A.~V.}\ \bibnamefont
  {Chubukov}}, \ and\ \bibinfo {author} {\bibfnamefont {J.}~\bibnamefont
  {Schmalian}},\ }\href@noop {} {\bibfield  {journal} {\bibinfo  {journal}
  {Nature Phys.}\ }\textbf {\bibinfo {volume} {10}},\ \bibinfo {pages} {97}
  (\bibinfo {year} {2014})}\BibitemShut {NoStop}%
\bibitem [{\citenamefont {Baek}\ \emph {et~al.}(2015)\citenamefont {Baek},
  \citenamefont {Efremov}, \citenamefont {Ok}, \citenamefont {Kim},
  \citenamefont {van~den Brink},\ and\ \citenamefont
  {B\"{u}chner}}]{BaekNatMater2016}%
  \BibitemOpen
  \bibfield  {author} {\bibinfo {author} {\bibfnamefont {S.-H.}\ \bibnamefont
  {Baek}}, \bibinfo {author} {\bibfnamefont {D.~V.}\ \bibnamefont {Efremov}},
  \bibinfo {author} {\bibfnamefont {J.~M.}\ \bibnamefont {Ok}}, \bibinfo
  {author} {\bibfnamefont {J.~S.}\ \bibnamefont {Kim}}, \bibinfo {author}
  {\bibfnamefont {J.}~\bibnamefont {van~den Brink}}, \ and\ \bibinfo {author}
  {\bibfnamefont {B.}~\bibnamefont {B\"{u}chner}},\ }\href {\doibase
  10.1038/nmat4138} {\bibfield  {journal} {\bibinfo  {journal} {Nature Mater.}\
  }\textbf {\bibinfo {volume} {14}},\ \bibinfo {pages} {210} (\bibinfo {year}
  {2015})}\BibitemShut {NoStop}%
\bibitem [{\citenamefont {Wang}\ \emph {et~al.}(2015)\citenamefont {Wang},
  \citenamefont {Kivelson},\ and\ \citenamefont {Lee}}]{WangFaNatPhys2015}%
  \BibitemOpen
  \bibfield  {author} {\bibinfo {author} {\bibfnamefont {F.}~\bibnamefont
  {Wang}}, \bibinfo {author} {\bibfnamefont {S.~A.}\ \bibnamefont {Kivelson}},
  \ and\ \bibinfo {author} {\bibfnamefont {D.-H.}\ \bibnamefont {Lee}},\
  }\href@noop {} {\bibfield  {journal} {\bibinfo  {journal} {Nature Phys.}\
  }\textbf {\bibinfo {volume} {11}},\ \bibinfo {pages} {959} (\bibinfo {year}
  {2015})}\BibitemShut {NoStop}%
\bibitem [{\citenamefont {Glasbrenner}\ \emph {et~al.}(2015)\citenamefont
  {Glasbrenner}, \citenamefont {Mazin}, \citenamefont {Jeschke}, \citenamefont
  {Hirschfeld}, \citenamefont {Fernandes},\ and\ \citenamefont
  {Valent\'{i}}}]{GlasbrenerNatPhys2015}%
  \BibitemOpen
  \bibfield  {author} {\bibinfo {author} {\bibfnamefont {J.~K.}\ \bibnamefont
  {Glasbrenner}}, \bibinfo {author} {\bibfnamefont {I.~I.}\ \bibnamefont
  {Mazin}}, \bibinfo {author} {\bibfnamefont {H.~O.}\ \bibnamefont {Jeschke}},
  \bibinfo {author} {\bibfnamefont {P.~J.}\ \bibnamefont {Hirschfeld}},
  \bibinfo {author} {\bibfnamefont {R.~M.}\ \bibnamefont {Fernandes}}, \ and\
  \bibinfo {author} {\bibfnamefont {R.}~\bibnamefont {Valent\'{i}}},\
  }\href@noop {} {\bibfield  {journal} {\bibinfo  {journal} {Nature Phys.}\
  }\textbf {\bibinfo {volume} {11}},\ \bibinfo {pages} {953} (\bibinfo {year}
  {2015})}\BibitemShut {NoStop}%
\bibitem [{\citenamefont {Yu}\ and\ \citenamefont
  {Si}(2015)}]{YuRongPhysRevLett2015}%
  \BibitemOpen
  \bibfield  {author} {\bibinfo {author} {\bibfnamefont {R.}~\bibnamefont
  {Yu}}\ and\ \bibinfo {author} {\bibfnamefont {Q.}~\bibnamefont {Si}},\ }\href
  {\doibase 10.1103/PhysRevLett.115.116401} {\bibfield  {journal} {\bibinfo
  {journal} {Phys. Rev. Lett.}\ }\textbf {\bibinfo {volume} {115}},\ \bibinfo
  {pages} {116401} (\bibinfo {year} {2015})}\BibitemShut {NoStop}%
\bibitem [{\citenamefont {Yamakawa}\ \emph {et~al.}(2016)\citenamefont
  {Yamakawa}, \citenamefont {Onari},\ and\ \citenamefont
  {Kontani}}]{YamakawaPhysRevX2016}%
  \BibitemOpen
  \bibfield  {author} {\bibinfo {author} {\bibfnamefont {Y.}~\bibnamefont
  {Yamakawa}}, \bibinfo {author} {\bibfnamefont {S.}~\bibnamefont {Onari}}, \
  and\ \bibinfo {author} {\bibfnamefont {H.}~\bibnamefont {Kontani}},\ }\href
  {\doibase 10.1103/PhysRevX.6.021032} {\bibfield  {journal} {\bibinfo
  {journal} {Phys. Rev. X}\ }\textbf {\bibinfo {volume} {6}},\ \bibinfo {pages}
  {021032} (\bibinfo {year} {2016})}\BibitemShut {NoStop}%
\bibitem [{\citenamefont {Watson}\ \emph {et~al.}(2015)\citenamefont {Watson},
  \citenamefont {Kim}, \citenamefont {Haghighirad}, \citenamefont {Davies},
  \citenamefont {McCollam}, \citenamefont {Narayanan}, \citenamefont {Blake},
  \citenamefont {Chen}, \citenamefont {Ghannadzadeh}, \citenamefont
  {Schofield}, \citenamefont {Hoesch}, \citenamefont {Meingast}, \citenamefont
  {Wolf},\ and\ \citenamefont {Coldea}}]{WatsonPhysRevB2015}%
  \BibitemOpen
  \bibfield  {author} {\bibinfo {author} {\bibfnamefont {M.~D.}\ \bibnamefont
  {Watson}}, \bibinfo {author} {\bibfnamefont {T.~K.}\ \bibnamefont {Kim}},
  \bibinfo {author} {\bibfnamefont {A.~A.}\ \bibnamefont {Haghighirad}},
  \bibinfo {author} {\bibfnamefont {N.~R.}\ \bibnamefont {Davies}}, \bibinfo
  {author} {\bibfnamefont {A.}~\bibnamefont {McCollam}}, \bibinfo {author}
  {\bibfnamefont {A.}~\bibnamefont {Narayanan}}, \bibinfo {author}
  {\bibfnamefont {S.~F.}\ \bibnamefont {Blake}}, \bibinfo {author}
  {\bibfnamefont {Y.~L.}\ \bibnamefont {Chen}}, \bibinfo {author}
  {\bibfnamefont {S.}~\bibnamefont {Ghannadzadeh}}, \bibinfo {author}
  {\bibfnamefont {A.~J.}\ \bibnamefont {Schofield}}, \bibinfo {author}
  {\bibfnamefont {M.}~\bibnamefont {Hoesch}}, \bibinfo {author} {\bibfnamefont
  {C.}~\bibnamefont {Meingast}}, \bibinfo {author} {\bibfnamefont
  {T.}~\bibnamefont {Wolf}}, \ and\ \bibinfo {author} {\bibfnamefont {A.~I.}\
  \bibnamefont {Coldea}},\ }\href {\doibase 10.1103/PhysRevB.91.155106}
  {\bibfield  {journal} {\bibinfo  {journal} {Phys. Rev. B}\ }\textbf {\bibinfo
  {volume} {91}},\ \bibinfo {pages} {155106} (\bibinfo {year}
  {2015})}\BibitemShut {NoStop}%
\bibitem [{\citenamefont {Johnson}\ \emph {et~al.}(2015)\citenamefont
  {Johnson}, \citenamefont {Yang}, \citenamefont {Rameau}, \citenamefont {Gu},
  \citenamefont {Pan}, \citenamefont {Valla}, \citenamefont {Weinert},\ and\
  \citenamefont {Fedorov}}]{JohnsonPhysRevLett2015}%
  \BibitemOpen
  \bibfield  {author} {\bibinfo {author} {\bibfnamefont {P.~D.}\ \bibnamefont
  {Johnson}}, \bibinfo {author} {\bibfnamefont {H.-B.}\ \bibnamefont {Yang}},
  \bibinfo {author} {\bibfnamefont {J.~D.}\ \bibnamefont {Rameau}}, \bibinfo
  {author} {\bibfnamefont {G.~D.}\ \bibnamefont {Gu}}, \bibinfo {author}
  {\bibfnamefont {Z.-H.}\ \bibnamefont {Pan}}, \bibinfo {author} {\bibfnamefont
  {T.}~\bibnamefont {Valla}}, \bibinfo {author} {\bibfnamefont
  {M.}~\bibnamefont {Weinert}}, \ and\ \bibinfo {author} {\bibfnamefont
  {A.~V.}\ \bibnamefont {Fedorov}},\ }\href {\doibase
  10.1103/PhysRevLett.114.167001} {\bibfield  {journal} {\bibinfo  {journal}
  {Phys. Rev. Lett.}\ }\textbf {\bibinfo {volume} {114}},\ \bibinfo {pages}
  {167001} (\bibinfo {year} {2015})}\BibitemShut {NoStop}%
\bibitem [{\citenamefont {Borisenko}\ \emph {et~al.}(2016)\citenamefont
  {Borisenko}, \citenamefont {Evtushinsky}, \citenamefont {Liu}, \citenamefont
  {Morozov}, \citenamefont {Kappenberger}, \citenamefont {Wurmehl},
  \citenamefont {B\"{u}chner}, \citenamefont {Yaresko}, \citenamefont {Kim},
  \citenamefont {Hoesch}, \citenamefont {Wolf},\ and\ \citenamefont
  {Zhigadlo}}]{BorisenkoNatPhys2015}%
  \BibitemOpen
  \bibfield  {author} {\bibinfo {author} {\bibfnamefont {S.~V.}\ \bibnamefont
  {Borisenko}}, \bibinfo {author} {\bibfnamefont {D.~V.}\ \bibnamefont
  {Evtushinsky}}, \bibinfo {author} {\bibfnamefont {Z.-H.}\ \bibnamefont
  {Liu}}, \bibinfo {author} {\bibfnamefont {I.}~\bibnamefont {Morozov}},
  \bibinfo {author} {\bibfnamefont {R.}~\bibnamefont {Kappenberger}}, \bibinfo
  {author} {\bibfnamefont {S.}~\bibnamefont {Wurmehl}}, \bibinfo {author}
  {\bibfnamefont {B.}~\bibnamefont {B\"{u}chner}}, \bibinfo {author}
  {\bibfnamefont {A.~N.}\ \bibnamefont {Yaresko}}, \bibinfo {author}
  {\bibfnamefont {T.~K.}\ \bibnamefont {Kim}}, \bibinfo {author} {\bibfnamefont
  {M.}~\bibnamefont {Hoesch}}, \bibinfo {author} {\bibfnamefont
  {T.}~\bibnamefont {Wolf}}, \ and\ \bibinfo {author} {\bibfnamefont {N.~D.}\
  \bibnamefont {Zhigadlo}},\ }\href@noop {} {\bibfield  {journal} {\bibinfo
  {journal} {Nature Phys.}\ }\textbf {\bibinfo {volume} {12}},\ \bibinfo
  {pages} {311} (\bibinfo {year} {2016})}\BibitemShut {NoStop}%
\bibitem [{\citenamefont {Dai}(2015)}]{DaiRevModPhys2015}%
  \BibitemOpen
  \bibfield  {author} {\bibinfo {author} {\bibfnamefont {P.}~\bibnamefont
  {Dai}},\ }\href {\doibase 10.1103/RevModPhys.87.855} {\bibfield  {journal}
  {\bibinfo  {journal} {Rev. Mod. Phys.}\ }\textbf {\bibinfo {volume} {87}},\
  \bibinfo {pages} {855} (\bibinfo {year} {2015})}\BibitemShut {NoStop}%
\bibitem [{\citenamefont {Park}\ \emph {et~al.}(2012)\citenamefont {Park},
  \citenamefont {Friemel}, \citenamefont {Loew}, \citenamefont {Hinkov},
  \citenamefont {Li}, \citenamefont {Min}, \citenamefont {Sun}, \citenamefont
  {Ivanov}, \citenamefont {Piovano}, \citenamefont {Lin}, \citenamefont
  {Keimer}, \citenamefont {Kwon},\ and\ \citenamefont
  {Inosov}}]{ParKPhysRevB2012}%
  \BibitemOpen
  \bibfield  {author} {\bibinfo {author} {\bibfnamefont {J.~T.}\ \bibnamefont
  {Park}}, \bibinfo {author} {\bibfnamefont {G.}~\bibnamefont {Friemel}},
  \bibinfo {author} {\bibfnamefont {T.}~\bibnamefont {Loew}}, \bibinfo {author}
  {\bibfnamefont {V.}~\bibnamefont {Hinkov}}, \bibinfo {author} {\bibfnamefont
  {Y.}~\bibnamefont {Li}}, \bibinfo {author} {\bibfnamefont {B.~H.}\
  \bibnamefont {Min}}, \bibinfo {author} {\bibfnamefont {D.~L.}\ \bibnamefont
  {Sun}}, \bibinfo {author} {\bibfnamefont {A.}~\bibnamefont {Ivanov}},
  \bibinfo {author} {\bibfnamefont {A.}~\bibnamefont {Piovano}}, \bibinfo
  {author} {\bibfnamefont {C.~T.}\ \bibnamefont {Lin}}, \bibinfo {author}
  {\bibfnamefont {B.}~\bibnamefont {Keimer}}, \bibinfo {author} {\bibfnamefont
  {Y.~S.}\ \bibnamefont {Kwon}}, \ and\ \bibinfo {author} {\bibfnamefont
  {D.~S.}\ \bibnamefont {Inosov}},\ }\href {\doibase
  10.1103/PhysRevB.86.024437} {\bibfield  {journal} {\bibinfo  {journal} {Phys.
  Rev. B}\ }\textbf {\bibinfo {volume} {86}},\ \bibinfo {pages} {024437}
  (\bibinfo {year} {2012})}\BibitemShut {NoStop}%
\bibitem [{\citenamefont {Qureshi}\ \emph {et~al.}(2012)\citenamefont
  {Qureshi}, \citenamefont {Steffens}, \citenamefont {Wurmehl}, \citenamefont
  {Aswartham}, \citenamefont {B\"uchner},\ and\ \citenamefont
  {Braden}}]{QureshiPhysRevB2012}%
  \BibitemOpen
  \bibfield  {author} {\bibinfo {author} {\bibfnamefont {N.}~\bibnamefont
  {Qureshi}}, \bibinfo {author} {\bibfnamefont {P.}~\bibnamefont {Steffens}},
  \bibinfo {author} {\bibfnamefont {S.}~\bibnamefont {Wurmehl}}, \bibinfo
  {author} {\bibfnamefont {S.}~\bibnamefont {Aswartham}}, \bibinfo {author}
  {\bibfnamefont {B.}~\bibnamefont {B\"uchner}}, \ and\ \bibinfo {author}
  {\bibfnamefont {M.}~\bibnamefont {Braden}},\ }\href {\doibase
  10.1103/PhysRevB.86.060410} {\bibfield  {journal} {\bibinfo  {journal} {Phys.
  Rev. B}\ }\textbf {\bibinfo {volume} {86}},\ \bibinfo {pages} {060410}
  (\bibinfo {year} {2012})}\BibitemShut {NoStop}%
\bibitem [{\citenamefont {Wang}\ \emph {et~al.}(2013)\citenamefont {Wang},
  \citenamefont {Zhang}, \citenamefont {Wang}, \citenamefont {Luo},
  \citenamefont {Regnault}, \citenamefont {Dai},\ and\ \citenamefont
  {Li}}]{WangCPhysRevX2013}%
  \BibitemOpen
  \bibfield  {author} {\bibinfo {author} {\bibfnamefont {C.}~\bibnamefont
  {Wang}}, \bibinfo {author} {\bibfnamefont {R.}~\bibnamefont {Zhang}},
  \bibinfo {author} {\bibfnamefont {F.}~\bibnamefont {Wang}}, \bibinfo {author}
  {\bibfnamefont {H.}~\bibnamefont {Luo}}, \bibinfo {author} {\bibfnamefont
  {L.~P.}\ \bibnamefont {Regnault}}, \bibinfo {author} {\bibfnamefont
  {P.}~\bibnamefont {Dai}}, \ and\ \bibinfo {author} {\bibfnamefont
  {Y.}~\bibnamefont {Li}},\ }\href {\doibase 10.1103/PhysRevX.3.041036}
  {\bibfield  {journal} {\bibinfo  {journal} {Phys. Rev. X}\ }\textbf {\bibinfo
  {volume} {3}},\ \bibinfo {pages} {041036} (\bibinfo {year}
  {2013})}\BibitemShut {NoStop}%
\bibitem [{\citenamefont {McQueen}\ \emph {et~al.}(2009)\citenamefont
  {McQueen}, \citenamefont {Williams}, \citenamefont {Stephens}, \citenamefont
  {Tao}, \citenamefont {Zhu}, \citenamefont {Ksenofontov}, \citenamefont
  {Casper}, \citenamefont {Felser},\ and\ \citenamefont
  {Cava}}]{McQueenPhysRevLett2009}%
  \BibitemOpen
  \bibfield  {author} {\bibinfo {author} {\bibfnamefont {T.~M.}\ \bibnamefont
  {McQueen}}, \bibinfo {author} {\bibfnamefont {A.~J.}\ \bibnamefont
  {Williams}}, \bibinfo {author} {\bibfnamefont {P.~W.}\ \bibnamefont
  {Stephens}}, \bibinfo {author} {\bibfnamefont {J.}~\bibnamefont {Tao}},
  \bibinfo {author} {\bibfnamefont {Y.}~\bibnamefont {Zhu}}, \bibinfo {author}
  {\bibfnamefont {V.}~\bibnamefont {Ksenofontov}}, \bibinfo {author}
  {\bibfnamefont {F.}~\bibnamefont {Casper}}, \bibinfo {author} {\bibfnamefont
  {C.}~\bibnamefont {Felser}}, \ and\ \bibinfo {author} {\bibfnamefont {R.~J.}\
  \bibnamefont {Cava}},\ }\href {\doibase 10.1103/PhysRevLett.103.057002}
  {\bibfield  {journal} {\bibinfo  {journal} {Phys. Rev. Lett.}\ }\textbf
  {\bibinfo {volume} {103}},\ \bibinfo {pages} {057002} (\bibinfo {year}
  {2009})}\BibitemShut {NoStop}%
\bibitem [{\citenamefont {Ye}\ \emph {et~al.}(2015)\citenamefont {Ye},
  \citenamefont {Zhang}, \citenamefont {Ning}, \citenamefont {Li},
  \citenamefont {Chen}, \citenamefont {Jia}, \citenamefont {Hashimoto},
  \citenamefont {Lu}, \citenamefont {Shen},\ and\ \citenamefont
  {Zhang}}]{YeZR2015}%
  \BibitemOpen
  \bibfield  {author} {\bibinfo {author} {\bibfnamefont {Z.~R.}\ \bibnamefont
  {Ye}}, \bibinfo {author} {\bibfnamefont {C.~F.}\ \bibnamefont {Zhang}},
  \bibinfo {author} {\bibfnamefont {H.~L.}\ \bibnamefont {Ning}}, \bibinfo
  {author} {\bibfnamefont {W.}~\bibnamefont {Li}}, \bibinfo {author}
  {\bibfnamefont {L.}~\bibnamefont {Chen}}, \bibinfo {author} {\bibfnamefont
  {T.}~\bibnamefont {Jia}}, \bibinfo {author} {\bibfnamefont {M.}~\bibnamefont
  {Hashimoto}}, \bibinfo {author} {\bibfnamefont {D.~H.}\ \bibnamefont {Lu}},
  \bibinfo {author} {\bibfnamefont {Z.-X.}\ \bibnamefont {Shen}}, \ and\
  \bibinfo {author} {\bibfnamefont {Y.}~\bibnamefont {Zhang}},\ }\href@noop {}
  {\enquote {\bibinfo {title} {{Simultaneous emergence of superconductivity,
  inter-pocket scattering and nematic fluctuation in potassium-coated FeSe
  superconductor}},}\ } (\bibinfo {year} {2015}),\ \bibinfo {note}
  {arXiv:1512.02526}\BibitemShut {NoStop}%
\bibitem [{\citenamefont {Rahn}\ \emph {et~al.}(2015)\citenamefont {Rahn},
  \citenamefont {Ewings}, \citenamefont {Sedlmaier}, \citenamefont {Clarke},\
  and\ \citenamefont {Boothroyd}}]{RahnPhysRevB2015}%
  \BibitemOpen
  \bibfield  {author} {\bibinfo {author} {\bibfnamefont {M.~C.}\ \bibnamefont
  {Rahn}}, \bibinfo {author} {\bibfnamefont {R.~A.}\ \bibnamefont {Ewings}},
  \bibinfo {author} {\bibfnamefont {S.~J.}\ \bibnamefont {Sedlmaier}}, \bibinfo
  {author} {\bibfnamefont {S.~J.}\ \bibnamefont {Clarke}}, \ and\ \bibinfo
  {author} {\bibfnamefont {A.~T.}\ \bibnamefont {Boothroyd}},\ }\href {\doibase
  10.1103/PhysRevB.91.180501} {\bibfield  {journal} {\bibinfo  {journal} {Phys.
  Rev. B}\ }\textbf {\bibinfo {volume} {91}},\ \bibinfo {pages} {180501}
  (\bibinfo {year} {2015})}\BibitemShut {NoStop}%
\bibitem [{\citenamefont {Steffens}\ \emph {et~al.}(2013)\citenamefont
  {Steffens}, \citenamefont {Lee}, \citenamefont {Qureshi}, \citenamefont
  {Kihou}, \citenamefont {Iyo}, \citenamefont {Eisaki},\ and\ \citenamefont
  {Braden}}]{SteffensPhysRevLett2013}%
  \BibitemOpen
  \bibfield  {author} {\bibinfo {author} {\bibfnamefont {P.}~\bibnamefont
  {Steffens}}, \bibinfo {author} {\bibfnamefont {C.~H.}\ \bibnamefont {Lee}},
  \bibinfo {author} {\bibfnamefont {N.}~\bibnamefont {Qureshi}}, \bibinfo
  {author} {\bibfnamefont {K.}~\bibnamefont {Kihou}}, \bibinfo {author}
  {\bibfnamefont {A.}~\bibnamefont {Iyo}}, \bibinfo {author} {\bibfnamefont
  {H.}~\bibnamefont {Eisaki}}, \ and\ \bibinfo {author} {\bibfnamefont
  {M.}~\bibnamefont {Braden}},\ }\href {\doibase
  10.1103/PhysRevLett.110.137001} {\bibfield  {journal} {\bibinfo  {journal}
  {Phys. Rev. Lett.}\ }\textbf {\bibinfo {volume} {110}},\ \bibinfo {pages}
  {137001} (\bibinfo {year} {2013})}\BibitemShut {NoStop}%
\bibitem [{\citenamefont {Zhang}\ \emph {et~al.}(2013)\citenamefont {Zhang},
  \citenamefont {Liu}, \citenamefont {Su}, \citenamefont {Regnault},
  \citenamefont {Wang}, \citenamefont {Tan}, \citenamefont {Br\"uckel},
  \citenamefont {Egami},\ and\ \citenamefont {Dai}}]{ZhangPhysRevB2013}%
  \BibitemOpen
  \bibfield  {author} {\bibinfo {author} {\bibfnamefont {C.}~\bibnamefont
  {Zhang}}, \bibinfo {author} {\bibfnamefont {M.}~\bibnamefont {Liu}}, \bibinfo
  {author} {\bibfnamefont {Y.}~\bibnamefont {Su}}, \bibinfo {author}
  {\bibfnamefont {L.-P.}\ \bibnamefont {Regnault}}, \bibinfo {author}
  {\bibfnamefont {M.}~\bibnamefont {Wang}}, \bibinfo {author} {\bibfnamefont
  {G.}~\bibnamefont {Tan}}, \bibinfo {author} {\bibfnamefont {T.}~\bibnamefont
  {Br\"uckel}}, \bibinfo {author} {\bibfnamefont {T.}~\bibnamefont {Egami}}, \
  and\ \bibinfo {author} {\bibfnamefont {P.}~\bibnamefont {Dai}},\ }\href
  {\doibase 10.1103/PhysRevB.87.081101} {\bibfield  {journal} {\bibinfo
  {journal} {Phys. Rev. B}\ }\textbf {\bibinfo {volume} {87}},\ \bibinfo
  {pages} {081101} (\bibinfo {year} {2013})}\BibitemShut {NoStop}%
\bibitem [{\citenamefont {Kasahara}\ \emph {et~al.}(2014)\citenamefont
  {Kasahara}, \citenamefont {Watashige}, \citenamefont {Hanaguri},
  \citenamefont {Kohsaka}, \citenamefont {Yamashita}, \citenamefont
  {Shimoyama}, \citenamefont {Mizukami}, \citenamefont {Endo}, \citenamefont
  {Ikeda}, \citenamefont {Aoyama}, \citenamefont {Terashima}, \citenamefont
  {Uji}, \citenamefont {Wolf}, \citenamefont {von L\"{o}hneysen}, \citenamefont
  {Shibauchi},\ and\ \citenamefont {Matsuda}}]{KasaharaPNAS2014}%
  \BibitemOpen
  \bibfield  {author} {\bibinfo {author} {\bibfnamefont {S.}~\bibnamefont
  {Kasahara}}, \bibinfo {author} {\bibfnamefont {T.}~\bibnamefont {Watashige}},
  \bibinfo {author} {\bibfnamefont {T.}~\bibnamefont {Hanaguri}}, \bibinfo
  {author} {\bibfnamefont {Y.}~\bibnamefont {Kohsaka}}, \bibinfo {author}
  {\bibfnamefont {T.}~\bibnamefont {Yamashita}}, \bibinfo {author}
  {\bibfnamefont {Y.}~\bibnamefont {Shimoyama}}, \bibinfo {author}
  {\bibfnamefont {Y.}~\bibnamefont {Mizukami}}, \bibinfo {author}
  {\bibfnamefont {R.}~\bibnamefont {Endo}}, \bibinfo {author} {\bibfnamefont
  {H.}~\bibnamefont {Ikeda}}, \bibinfo {author} {\bibfnamefont
  {K.}~\bibnamefont {Aoyama}}, \bibinfo {author} {\bibfnamefont
  {T.}~\bibnamefont {Terashima}}, \bibinfo {author} {\bibfnamefont
  {S.}~\bibnamefont {Uji}}, \bibinfo {author} {\bibfnamefont {T.}~\bibnamefont
  {Wolf}}, \bibinfo {author} {\bibfnamefont {H.}~\bibnamefont {von
  L\"{o}hneysen}}, \bibinfo {author} {\bibfnamefont {T.}~\bibnamefont
  {Shibauchi}}, \ and\ \bibinfo {author} {\bibfnamefont {Y.}~\bibnamefont
  {Matsuda}},\ }\href {\doibase 10.1073/pnas.1413477111} {\bibfield  {journal}
  {\bibinfo  {journal} {Proc. Natl Acad. Sci. USA}\ }\textbf {\bibinfo {volume}
  {111}},\ \bibinfo {pages} {16309} (\bibinfo {year} {2014})}\BibitemShut
  {NoStop}%
\bibitem [{\citenamefont {Lin}\ \emph {et~al.}(2011)\citenamefont {Lin},
  \citenamefont {Hsieh}, \citenamefont {Chareev}, \citenamefont {Vasiliev},
  \citenamefont {Parsons},\ and\ \citenamefont {Yang}}]{LinPhysRevB2011}%
  \BibitemOpen
  \bibfield  {author} {\bibinfo {author} {\bibfnamefont {J.-Y.}\ \bibnamefont
  {Lin}}, \bibinfo {author} {\bibfnamefont {Y.~S.}\ \bibnamefont {Hsieh}},
  \bibinfo {author} {\bibfnamefont {D.~A.}\ \bibnamefont {Chareev}}, \bibinfo
  {author} {\bibfnamefont {A.~N.}\ \bibnamefont {Vasiliev}}, \bibinfo {author}
  {\bibfnamefont {Y.}~\bibnamefont {Parsons}}, \ and\ \bibinfo {author}
  {\bibfnamefont {H.~D.}\ \bibnamefont {Yang}},\ }\href {\doibase
  10.1103/PhysRevB.84.220507} {\bibfield  {journal} {\bibinfo  {journal} {Phys.
  Rev. B}\ }\textbf {\bibinfo {volume} {84}},\ \bibinfo {pages} {220507}
  (\bibinfo {year} {2011})}\BibitemShut {NoStop}%
\bibitem [{\citenamefont {Braden}\ \emph {et~al.}(2004)\citenamefont {Braden},
  \citenamefont {Steffens}, \citenamefont {Sidis}, \citenamefont {Kulda},
  \citenamefont {Bourges}, \citenamefont {Hayden}, \citenamefont {Kikugawa},\
  and\ \citenamefont {Maeno}}]{BradenPhysRevLett2004}%
  \BibitemOpen
  \bibfield  {author} {\bibinfo {author} {\bibfnamefont {M.}~\bibnamefont
  {Braden}}, \bibinfo {author} {\bibfnamefont {P.}~\bibnamefont {Steffens}},
  \bibinfo {author} {\bibfnamefont {Y.}~\bibnamefont {Sidis}}, \bibinfo
  {author} {\bibfnamefont {J.}~\bibnamefont {Kulda}}, \bibinfo {author}
  {\bibfnamefont {P.}~\bibnamefont {Bourges}}, \bibinfo {author} {\bibfnamefont
  {S.}~\bibnamefont {Hayden}}, \bibinfo {author} {\bibfnamefont
  {N.}~\bibnamefont {Kikugawa}}, \ and\ \bibinfo {author} {\bibfnamefont
  {Y.}~\bibnamefont {Maeno}},\ }\href {\doibase 10.1103/PhysRevLett.92.097402}
  {\bibfield  {journal} {\bibinfo  {journal} {Phys. Rev. Lett.}\ }\textbf
  {\bibinfo {volume} {92}},\ \bibinfo {pages} {097402} (\bibinfo {year}
  {2004})}\BibitemShut {NoStop}%
\bibitem [{\citenamefont {Eremin}\ \emph {et~al.}(2002)\citenamefont {Eremin},
  \citenamefont {Manske},\ and\ \citenamefont
  {Bennemann}}]{EreminPhysRevB2002}%
  \BibitemOpen
  \bibfield  {author} {\bibinfo {author} {\bibfnamefont {I.}~\bibnamefont
  {Eremin}}, \bibinfo {author} {\bibfnamefont {D.}~\bibnamefont {Manske}}, \
  and\ \bibinfo {author} {\bibfnamefont {K.~H.}\ \bibnamefont {Bennemann}},\
  }\href {\doibase 10.1103/PhysRevB.65.220502} {\bibfield  {journal} {\bibinfo
  {journal} {Phys. Rev. B}\ }\textbf {\bibinfo {volume} {65}},\ \bibinfo
  {pages} {220502} (\bibinfo {year} {2002})}\BibitemShut {NoStop}%
\bibitem [{\citenamefont {Fedorov}\ \emph {et~al.}(2016)\citenamefont
  {Fedorov}, \citenamefont {Yaresko}, \citenamefont {Kim}, \citenamefont
  {Kushnirenko}, \citenamefont {Haubold}, \citenamefont {Wolf}, \citenamefont
  {Hoesch}, \citenamefont {Gr\"{u}neis}, \citenamefont {B\"{u}chner},\ and\
  \citenamefont {Borisenko}}]{Fedorov2016}%
  \BibitemOpen
  \bibfield  {author} {\bibinfo {author} {\bibfnamefont {A.}~\bibnamefont
  {Fedorov}}, \bibinfo {author} {\bibfnamefont {A.}~\bibnamefont {Yaresko}},
  \bibinfo {author} {\bibfnamefont {T.~K.}\ \bibnamefont {Kim}}, \bibinfo
  {author} {\bibfnamefont {E.}~\bibnamefont {Kushnirenko}}, \bibinfo {author}
  {\bibfnamefont {E.}~\bibnamefont {Haubold}}, \bibinfo {author} {\bibfnamefont
  {T.}~\bibnamefont {Wolf}}, \bibinfo {author} {\bibfnamefont {M.}~\bibnamefont
  {Hoesch}}, \bibinfo {author} {\bibfnamefont {A.}~\bibnamefont {Gr\"{u}neis}},
  \bibinfo {author} {\bibfnamefont {B.}~\bibnamefont {B\"{u}chner}}, \ and\
  \bibinfo {author} {\bibfnamefont {S.~V.}\ \bibnamefont {Borisenko}},\
  }\href@noop {} {\enquote {\bibinfo {title} {{Effect of nematic ordering on
  electronic structure of FeSe}},}\ } (\bibinfo {year} {2016}),\ \bibinfo
  {note} {arXiv:1606.03022}\BibitemShut {NoStop}%
\bibitem [{\citenamefont {Wang}\ \emph
  {et~al.}(2016{\natexlab{b}})\citenamefont {Wang}, \citenamefont {Sun},
  \citenamefont {Cui}, \citenamefont {Song}, \citenamefont {Li}, \citenamefont
  {Yu}, \citenamefont {Lei},\ and\ \citenamefont {Yu}}]{Wang2016}%
  \BibitemOpen
  \bibfield  {author} {\bibinfo {author} {\bibfnamefont {P.}~\bibnamefont
  {Wang}}, \bibinfo {author} {\bibfnamefont {S.}~\bibnamefont {Sun}}, \bibinfo
  {author} {\bibfnamefont {Y.}~\bibnamefont {Cui}}, \bibinfo {author}
  {\bibfnamefont {W.}~\bibnamefont {Song}}, \bibinfo {author} {\bibfnamefont
  {T.}~\bibnamefont {Li}}, \bibinfo {author} {\bibfnamefont {R.}~\bibnamefont
  {Yu}}, \bibinfo {author} {\bibfnamefont {H.}~\bibnamefont {Lei}}, \ and\
  \bibinfo {author} {\bibfnamefont {W.}~\bibnamefont {Yu}},\ }\href@noop {}
  {\enquote {\bibinfo {title} {{Unlocked stripe-order antiferromagnetism in
  FeSe under pressure}},}\ } (\bibinfo {year} {2016}{\natexlab{b}}),\ \bibinfo
  {note} {arXiv:1603.04589}\BibitemShut {NoStop}%
\bibitem [{\citenamefont {Wa\ss{}er}\ \emph {et~al.}(2015)\citenamefont
  {Wa\ss{}er}, \citenamefont {Schneidewind}, \citenamefont {Sidis},
  \citenamefont {Wurmehl}, \citenamefont {Aswartham}, \citenamefont
  {B\"uchner},\ and\ \citenamefont {Braden}}]{WassPhysRevB2015}%
  \BibitemOpen
  \bibfield  {author} {\bibinfo {author} {\bibfnamefont {F.}~\bibnamefont
  {Wa\ss{}er}}, \bibinfo {author} {\bibfnamefont {A.}~\bibnamefont
  {Schneidewind}}, \bibinfo {author} {\bibfnamefont {Y.}~\bibnamefont {Sidis}},
  \bibinfo {author} {\bibfnamefont {S.}~\bibnamefont {Wurmehl}}, \bibinfo
  {author} {\bibfnamefont {S.}~\bibnamefont {Aswartham}}, \bibinfo {author}
  {\bibfnamefont {B.}~\bibnamefont {B\"uchner}}, \ and\ \bibinfo {author}
  {\bibfnamefont {M.}~\bibnamefont {Braden}},\ }\href {\doibase
  10.1103/PhysRevB.91.060505} {\bibfield  {journal} {\bibinfo  {journal} {Phys.
  Rev. B}\ }\textbf {\bibinfo {volume} {91}},\ \bibinfo {pages} {060505}
  (\bibinfo {year} {2015})}\BibitemShut {NoStop}%
\bibitem [{\citenamefont {Allred}\ \emph {et~al.}(2016)\citenamefont {Allred},
  \citenamefont {Taddei}, \citenamefont {Bugaris}, \citenamefont {Krogstad},
  \citenamefont {Lapidus}, \citenamefont {Chung}, \citenamefont {Claus},
  \citenamefont {Kanatzidis}, \citenamefont {Brown}, \citenamefont {Kang},
  \citenamefont {Fernandes}, \citenamefont {Eremin}, \citenamefont
  {Rosenkranz}, \citenamefont {Chmaissem},\ and\ \citenamefont
  {Osborn}}]{AllredNatPhys2016}%
  \BibitemOpen
  \bibfield  {author} {\bibinfo {author} {\bibfnamefont {J.~M.}\ \bibnamefont
  {Allred}}, \bibinfo {author} {\bibfnamefont {K.~M.}\ \bibnamefont {Taddei}},
  \bibinfo {author} {\bibfnamefont {D.~E.}\ \bibnamefont {Bugaris}}, \bibinfo
  {author} {\bibfnamefont {M.~J.}\ \bibnamefont {Krogstad}}, \bibinfo {author}
  {\bibfnamefont {S.~H.}\ \bibnamefont {Lapidus}}, \bibinfo {author}
  {\bibfnamefont {D.~Y.}\ \bibnamefont {Chung}}, \bibinfo {author}
  {\bibfnamefont {H.}~\bibnamefont {Claus}}, \bibinfo {author} {\bibfnamefont
  {M.~G.}\ \bibnamefont {Kanatzidis}}, \bibinfo {author} {\bibfnamefont
  {D.~E.}\ \bibnamefont {Brown}}, \bibinfo {author} {\bibfnamefont
  {J.}~\bibnamefont {Kang}}, \bibinfo {author} {\bibfnamefont {R.~M.}\
  \bibnamefont {Fernandes}}, \bibinfo {author} {\bibfnamefont {I.}~\bibnamefont
  {Eremin}}, \bibinfo {author} {\bibfnamefont {S.}~\bibnamefont {Rosenkranz}},
  \bibinfo {author} {\bibfnamefont {O.}~\bibnamefont {Chmaissem}}, \ and\
  \bibinfo {author} {\bibfnamefont {R.}~\bibnamefont {Osborn}},\ }\href@noop {}
  {\bibfield  {journal} {\bibinfo  {journal} {Nature Phys.}\ }\textbf {\bibinfo
  {volume} {12}},\ \bibinfo {pages} {493} (\bibinfo {year} {2016})}\BibitemShut
  {NoStop}%
\bibitem [{\citenamefont {Christensen}\ \emph {et~al.}(2015)\citenamefont
  {Christensen}, \citenamefont {Kang}, \citenamefont {Andersen}, \citenamefont
  {Eremin},\ and\ \citenamefont {Fernandes}}]{ChristensenPhysRevB2015}%
  \BibitemOpen
  \bibfield  {author} {\bibinfo {author} {\bibfnamefont {M.~H.}\ \bibnamefont
  {Christensen}}, \bibinfo {author} {\bibfnamefont {J.}~\bibnamefont {Kang}},
  \bibinfo {author} {\bibfnamefont {B.~M.}\ \bibnamefont {Andersen}}, \bibinfo
  {author} {\bibfnamefont {I.}~\bibnamefont {Eremin}}, \ and\ \bibinfo {author}
  {\bibfnamefont {R.~M.}\ \bibnamefont {Fernandes}},\ }\href {\doibase
  10.1103/PhysRevB.92.214509} {\bibfield  {journal} {\bibinfo  {journal} {Phys.
  Rev. B}\ }\textbf {\bibinfo {volume} {92}},\ \bibinfo {pages} {214509}
  (\bibinfo {year} {2015})}\BibitemShut {NoStop}%
\bibitem [{\citenamefont {Yuan}\ \emph {et~al.}(2014)\citenamefont {Yuan},
  \citenamefont {Mak},\ and\ \citenamefont {Law}}]{YuanPhysRevLett2014}%
  \BibitemOpen
  \bibfield  {author} {\bibinfo {author} {\bibfnamefont {N.~F.~Q.}\
  \bibnamefont {Yuan}}, \bibinfo {author} {\bibfnamefont {K.~F.}\ \bibnamefont
  {Mak}}, \ and\ \bibinfo {author} {\bibfnamefont {K.~T.}\ \bibnamefont
  {Law}},\ }\href {\doibase 10.1103/PhysRevLett.113.097001} {\bibfield
  {journal} {\bibinfo  {journal} {Phys. Rev. Lett.}\ }\textbf {\bibinfo
  {volume} {113}},\ \bibinfo {pages} {097001} (\bibinfo {year}
  {2014})}\BibitemShut {NoStop}%
\bibitem [{\citenamefont {Lu}\ \emph {et~al.}(2015)\citenamefont {Lu},
  \citenamefont {Zheliuk}, \citenamefont {Leermakers}, \citenamefont {Yuan},
  \citenamefont {Zeitler}, \citenamefont {Law},\ and\ \citenamefont
  {Ye}}]{LuScience2015}%
  \BibitemOpen
  \bibfield  {author} {\bibinfo {author} {\bibfnamefont {J.~M.}\ \bibnamefont
  {Lu}}, \bibinfo {author} {\bibfnamefont {O.}~\bibnamefont {Zheliuk}},
  \bibinfo {author} {\bibfnamefont {I.}~\bibnamefont {Leermakers}}, \bibinfo
  {author} {\bibfnamefont {N.~F.~Q.}\ \bibnamefont {Yuan}}, \bibinfo {author}
  {\bibfnamefont {U.}~\bibnamefont {Zeitler}}, \bibinfo {author} {\bibfnamefont
  {K.~T.}\ \bibnamefont {Law}}, \ and\ \bibinfo {author} {\bibfnamefont
  {J.~T.}\ \bibnamefont {Ye}},\ }\href {\doibase 10.1126/science.aab2277}
  {\bibfield  {journal} {\bibinfo  {journal} {Science}\ }\textbf {\bibinfo
  {volume} {350}},\ \bibinfo {pages} {1353} (\bibinfo {year}
  {2015})}\BibitemShut {NoStop}%
\bibitem [{\citenamefont {Her}\ \emph {et~al.}(2015)\citenamefont {Her},
  \citenamefont {Kohama}, \citenamefont {Matsuda}, \citenamefont {Kindo},
  \citenamefont {Yang}, \citenamefont {Chareev}, \citenamefont {Mitrofanova},
  \citenamefont {Volkova}, \citenamefont {Vasiliev},\ and\ \citenamefont
  {Lin}}]{HerSupercondSciTech2011}%
  \BibitemOpen
  \bibfield  {author} {\bibinfo {author} {\bibfnamefont {J.~L.}\ \bibnamefont
  {Her}}, \bibinfo {author} {\bibfnamefont {Y.}~\bibnamefont {Kohama}},
  \bibinfo {author} {\bibfnamefont {Y.~H.}\ \bibnamefont {Matsuda}}, \bibinfo
  {author} {\bibfnamefont {K.}~\bibnamefont {Kindo}}, \bibinfo {author}
  {\bibfnamefont {W.-H.}\ \bibnamefont {Yang}}, \bibinfo {author}
  {\bibfnamefont {D.~A.}\ \bibnamefont {Chareev}}, \bibinfo {author}
  {\bibfnamefont {E.~S.}\ \bibnamefont {Mitrofanova}}, \bibinfo {author}
  {\bibfnamefont {O.~S.}\ \bibnamefont {Volkova}}, \bibinfo {author}
  {\bibfnamefont {A.~N.}\ \bibnamefont {Vasiliev}}, \ and\ \bibinfo {author}
  {\bibfnamefont {J.-Y.}\ \bibnamefont {Lin}},\ }\href@noop {} {\bibfield
  {journal} {\bibinfo  {journal} {Supercond. Sci. Technol.}\ }\textbf {\bibinfo
  {volume} {28}},\ \bibinfo {pages} {045013} (\bibinfo {year}
  {2015})}\BibitemShut {NoStop}%
\bibitem [{\citenamefont {Yin}\ \emph {et~al.}(2015)\citenamefont {Yin},
  \citenamefont {Wu}, \citenamefont {Wang}, \citenamefont {Ye}, \citenamefont
  {Gong}, \citenamefont {Hou}, \citenamefont {Shan}, \citenamefont {Li},
  \citenamefont {Liang}, \citenamefont {Wu}, \citenamefont {Li}, \citenamefont
  {Ting}, \citenamefont {Wang}, \citenamefont {Hu}, \citenamefont {Hor},
  \citenamefont {Ding},\ and\ \citenamefont {Pan}}]{YinNatPhys2015}%
  \BibitemOpen
  \bibfield  {author} {\bibinfo {author} {\bibfnamefont {J.-X.}\ \bibnamefont
  {Yin}}, \bibinfo {author} {\bibfnamefont {Z.}~\bibnamefont {Wu}}, \bibinfo
  {author} {\bibfnamefont {J.-H.}\ \bibnamefont {Wang}}, \bibinfo {author}
  {\bibfnamefont {Z.-Y.}\ \bibnamefont {Ye}}, \bibinfo {author} {\bibfnamefont
  {J.}~\bibnamefont {Gong}}, \bibinfo {author} {\bibfnamefont {X.-Y.}\
  \bibnamefont {Hou}}, \bibinfo {author} {\bibfnamefont {L.}~\bibnamefont
  {Shan}}, \bibinfo {author} {\bibfnamefont {A.}~\bibnamefont {Li}}, \bibinfo
  {author} {\bibfnamefont {X.-J.}\ \bibnamefont {Liang}}, \bibinfo {author}
  {\bibfnamefont {X.-X.}\ \bibnamefont {Wu}}, \bibinfo {author} {\bibfnamefont
  {J.}~\bibnamefont {Li}}, \bibinfo {author} {\bibfnamefont {C.-S.}\
  \bibnamefont {Ting}}, \bibinfo {author} {\bibfnamefont {Z.-Q.}\ \bibnamefont
  {Wang}}, \bibinfo {author} {\bibfnamefont {J.-P.}\ \bibnamefont {Hu}},
  \bibinfo {author} {\bibfnamefont {P.-H.}\ \bibnamefont {Hor}}, \bibinfo
  {author} {\bibfnamefont {H.}~\bibnamefont {Ding}}, \ and\ \bibinfo {author}
  {\bibfnamefont {S.~H.}\ \bibnamefont {Pan}},\ }\href@noop {} {\bibfield
  {journal} {\bibinfo  {journal} {Nature Phys.}\ }\textbf {\bibinfo {volume}
  {11}},\ \bibinfo {pages} {543} (\bibinfo {year} {2015})}\BibitemShut
  {NoStop}%
\end{thebibliography}%

\end{document}